\documentclass[english,showpacs,floatfix]{revtex4}
\usepackage[dvips]{graphicx}
\usepackage{longtable}
\usepackage{amsmath,amssymb}
\usepackage{dcolumn}

\usepackage{babel}
\usepackage[latin1]{inputenc}

\begin{document}

\title{The phase transition and the Quasi-Normal Modes of black Holes}

\author{Jianyong Shen}
\author{Bin Wang}
\email{wangb@fudan.edu.cn}
\affiliation{Department of Physics, Fudan University, 200433
Shanghai}

\author{Chi-Yong Lin}
\email{lcyong@mail.ndhu.edu.tw} \affiliation{Department of Physics,
National Dong Hwa University, Shoufeng, 974 Hualien}

\author{Rong-Geng Cai}
\email{cairg@itp.ac.cn} \affiliation{Institute of Theoretical
Physics, Chinese Academy of Sciences, P.O. Box 2735, 100080
Beijing}

\author{Ru-Keng Su}
\email{rksu@fudan.ac.cn}
\affiliation{China Center of Advanced Science and Technology
(World Laboratory) P.O. Box 8730, 100080
Beijing}\affiliation{Department of Physics, Fudan University,
200433 Shanghai}


\begin{abstract}
We reexamined the argument that the quasinormal modes could be a
probe of the phase transition of a topological black hole to a
hairy configuration by investigating general scalar perturbations.
We found further evidence in the quasinormal modes for this phase
transition. For the general black hole configurations, we observed
that although the quasinormal modes can present us different
phases of different configurations, there is no dramatic change in
the slope of quasinormal frequencies at the critical point of the
phase transition. More detailed studies of quasinormal modes are
needed to reveal the subtle behavior of the phase transition.

\end{abstract}

\maketitle

\section{Introduction}
Black holes' quasinormal modes (QNMs) have been an intriguing
subject of discussions in the past decades \cite{1,2,3}. The QNM is
believed as characteristic sound of black holes, which describes the
damped oscillations under perturbations in the surrounding geometry
of a black hole with frequencies and damping times of the
oscillations entirely fixed by the black hole parameters. The QNMs
of black holes have potential astrophysical interest since it could
lead to the direct identification of the black hole existence
through gravitational wave observation to be realized in the near
future\cite{1,2}. Despite the astrophysical interest, it has been
argued that the black holes' QNM could be a testing ground for
fundamental physics. Motivated by the discovery of the AdS/CFT
correspondence, the investigation of QNM in anti-de Sitter(AdS)
spacetimes became appealing in the past several years. It was argued
that the QNMs of AdS black holes have direct interpretation in term
of the dual conformal field theory(CFT)\cite{3,4,5,6,7,8,9}.
Attempts of using QNMs to investigate the dS/CFT correspondence have
also been given\cite{10}. Recently QNMs in asymptotically flat
spaces have acquired further attention, since the possible
connection between the classical vibrations of a black hole
spacetime and various quantum aspects was proposed by relating the
real part of the QNM frequencies to the Barbero-Immirzi(BI)
parameter, a factor introduced by hand in order that loop quantum
gravity reproduces correctly the black hole entropy \cite{11}. The
extension has been done in the dS background \cite{12}, however in
the AdS black hole spacetime, the direct relation has not been
found\cite{13}.

Recently further motivation of studying the QNMs has been pointed
out in \cite{14} by arguing that QNMs can reflect the black hole
phase transition. By calculating the simplest possible QNMs of
electromagnetic perturbations in the background of the MTZ black
hole obtained in \cite{15}, it was claimed in \cite{14} that they
found the evidence of the phase transition in the QNMs behavior
for small topological black holes with scalar hair. Further they
claimed that at the critical temperature, the continuously
matching of thermodynamical functions leads the phase transition
to be of the second order and they gave the order parameter of the
phase transition.  Their result is interesting, since it might be
the first phenomenon telling us the existence of the phase
transition in black hole physics.

Is the obtained signature of phase transition in the QNMs of
electromagnetic perturbation just an accident? Does the connection
between QNMs and phase transition hold for more general field
perturbations such as the general scalar and gravitational
perturbations? Can the QNMs be an effective probe of phase
transitions in more general black hole configurations? In this
paper we are trying to answer these questions. We will first
extend the study in \cite{14} by investigating the scalar
perturbations around the MTZ black holes and computing its
possible QNMs. MTZ black hole is an exact topological black hole
solution wearing minimally coupled nontrivial scalar field. The
study of the scalar perturbation in this black hole background is
interesting. Numerically we will show that the change of slope of
the QNMs again appears as we decrease the value of the horizon
radius below the critical value as that in the electromagnetic
perturbation \cite{14} which shows the phase transition of a
vacuum topological black hole to the MTZ black hole with scalar
hair.  To examine whether the QNM is an effective probe of the
phase transition in general configurations, we will calculate the
QNMs of scalar perturbation of the AdS black holes with Ricci flat
horizons using AdS soliton as the background. It was found that
there is a phase transition analogous to the Hawking-Page
transition for AdS black holes \cite{HM, surya}. We are going to
study whether the signature of this phase transition can be
reflected in the QNMs behavior.

\section{The topological black hole with scalar hair}
Considering the four-dimensional gravity with negative
cosmological constant ($\Lambda = -3/l^2$) and a scalar field
described by the action
\begin{equation}\label{e1}
I = \int {d^4 x\sqrt { - g} [\frac{{R + 6l^{ - 2} }}{16 \pi G} -
\frac{1}{2}\partial ^\mu  \phi \partial _\mu  \phi  - V(\phi )]}
\end{equation}
where the potential is given by $ V(\phi ) =  - \frac{3}{{4\pi
Gl^2 }}\sinh ^2 \sqrt {\frac{{4\pi G}}{3}} \phi$, we have the
equation of motion of gravitational field
\begin{equation}\label{e2}
G_{\mu \nu }  + \Lambda g_{\mu \nu }  = 8 \pi G(\nabla _\mu  \phi
\nabla _\nu  \phi  - \frac{1}{2}g_{\mu \nu } (\nabla \phi )^2  -
g_{\mu \nu } V(\phi ))
\end{equation}
and the scalar field satisfying
\begin{equation}\label{e3}
\nabla ^2 \phi  - \frac{{dV}}{{d\phi }} = 0.
\end{equation}
The exact solution of topological black hole with the scalar field
can be found with the metric \cite{15}
\begin{equation}\label{e4}
ds^2  = \frac{{r(r + 2G \mu )}}{{(r + G \mu )^2 }}[ - (\frac{{r^2
}}{{l^2 }} - (1 + \frac{G \mu }{r})^2 )dt^2  + (\frac{{r^2 }}{{l^2
}} - (1 + \frac{G \mu }{r})^2 )^{ - 1} dr^2  + r^2 d\sigma ^2 ]
\end{equation}
and the scalar field reads
\begin{equation}\label{e5}
\phi  = \sqrt {\frac {3}{{4 \pi G}}} \tanh ^{ - 1} \frac{G \mu
}{{r + G \mu }}.
\end{equation}
Here $d \sigma^2$ is the line element of the base manifold $\Sigma$,
which has negative constant curvature \cite{15}. The constant $\mu$
stands for the mass of black hole. The range of $r$ is taken as
$r>-2\mu$ for negative mass and $r>0$ otherwise to avoid the
singularities of the curvature and the scalar field, where the
conformal factor vanishes. The even horizon is given by
\begin{equation}\label{e6}
r_ +   = \frac{l}{2}(1 + \sqrt {1 + 4G \mu /l} ).
\end{equation}

From Eq.(\ref{e5}), we can see that the black hole always wears a
scalar field for fixed non-zero mass. When $\mu=0$, the metric
Eq.(\ref{e4}) stands for a locally AdS spacetime without the scalar
field
\begin{equation}\label{e7}
ds^2  =  - (\frac{{r^2 }}{{l^2 }} - 1)dt^2  + (\frac{{r^2 }}{{l^2 }}
- 1)^{ - 1} dr^2  + r^2 d\sigma ^2.
\end{equation}
The interesting fact is that the above AdS spacetime can also be
obtained from a topological black hole without scalar field
\begin{equation}\label{e8}
ds^2  =  - (\frac{{\rho ^2 }}{{l^2 }} - 1 - \frac{{2G \mu }}{\rho
})dt^2  + (\frac{{\rho ^2 }}{{l^2 }} - 1 - \frac{{2G \mu }}{\rho
})^{ - 1} d\rho ^2  + \rho ^2 d\sigma ^2,
\end{equation}
when the mass $\mu$ goes to zero. This means that for a given mass,
there are two branches of different black hole solutions. By
computing and comparing their free energies, Martinez et.al.
\cite{15} suggested that a second order phase transition exists when
the temperature (the even horizon) crosses a critical value $ T_c =
\frac{1}{{2\pi l}}$ ($ r_c = l $). When $T > T_c$ ($r_+ > l$), both
of black holes have positive mass. The black hole will absorb the
scalar field dress and turn into the stabler bare one Eq.(\ref{e8}).
When $T <T_c$ ($r_+ < l$), both of black holes have negative mass. A
process of dressing up scalar field will happen for the bare black
hole and the MTZ black hole has higher stability. Koutsoumbas {\it
et al.} \cite{14} further studied the behavior of the free energies
of both black holes and related it to the order parameter defined by
\begin{equation}\label{e9}
\lambda  = \left\{ \begin{array}{l}
 \frac{{T_0  - T}}{{T_0  + T}}, \quad T < T_0  \\
 0,~~~~~ \quad T > T_0 . \\
 \end{array} \right.
\end{equation}

\begin{figure}[ht]
\vspace*{0cm}
\begin{minipage}{0.3\textwidth}
\resizebox{1.1\linewidth}{!}{\includegraphics*{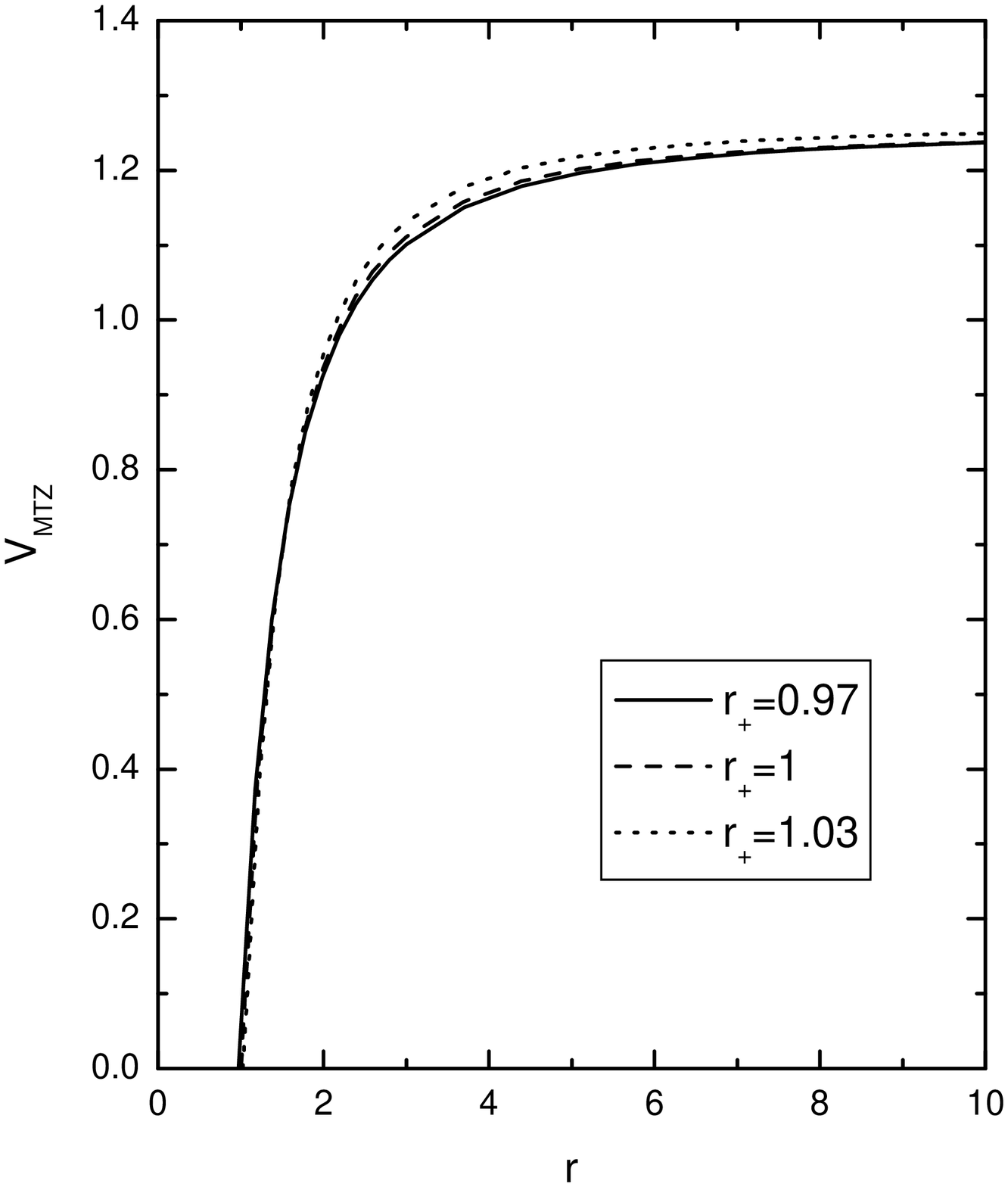}} \nonumber
\end{minipage}\nonumber
\begin{minipage}{0.3\textwidth}
\vspace*{0.0cm}
\resizebox{1.1\linewidth}{!}{\includegraphics*{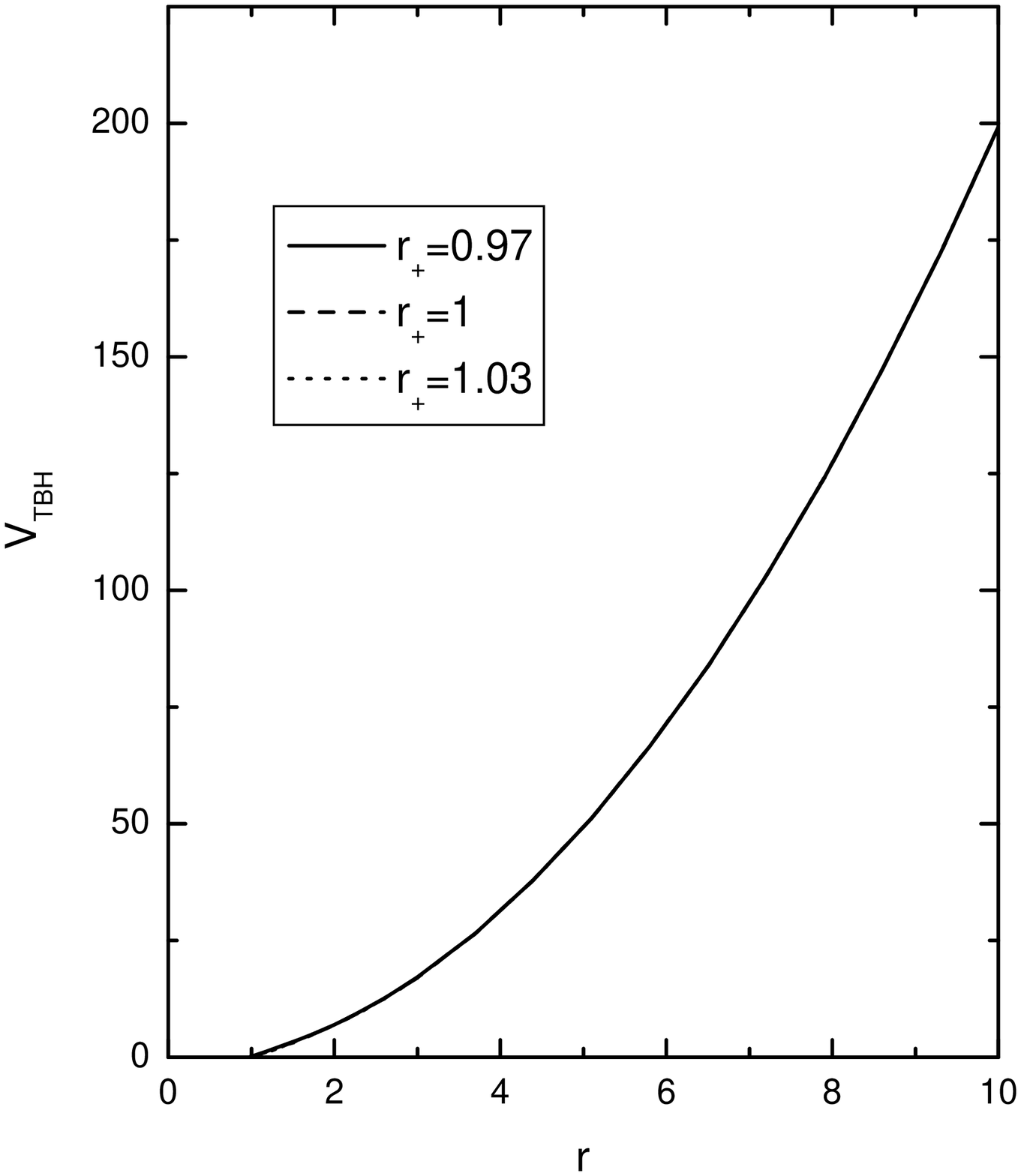}} \nonumber
\end{minipage} \nonumber
\caption{{The effective potentials of the scalar perturbations. The
left one stands for the MTZ black hole and the right for the
topological back hole, when $\xi = 1$ in the Eq.(\ref{e23}) and
Eq.(\ref{e26}).}} \label{f1}
\end{figure}

By computing the simplest possible QNMs of electromagnetic
perturbation in the MTZ and TBH backgrounds, Koutsoumbas {\it et
al.} \cite{14} showed that near the critical temperature and for
small black holes there is clear evidence in QNMs on the second
order phase transition between the vacuum topological black hole and
the MTZ black hole with scalar hair. It is of interest to generalize
their study to QNMs of more general fields perturbations, such as
the scalar field perturbation. In this subsection we calculate
numerically the QNMs of scalar perturbation for both MTZ and
topological black holes and examine the phase transition footprint.

Since the conformal factor in Eq.(\ref{e4}) does not play an
important role, for the convenience of discussion and calculation,
we make a conformal transformation to get rid of this factor and
take $8 \pi G=1$, $l=1$. In this frame the field equations are
\cite{winstanley}
\begin{equation}\label{e17}
(1 - \frac{1}{6}\phi ^2 )G_{\mu \nu }  - 3 g_{\mu \nu }
=\frac{2}{3}\nabla _\mu  \phi \nabla _\nu  \phi  - \frac{1}{6}g_{\mu
\nu } (\nabla \phi )^2  + \frac{1}{3}g_{\mu \nu } \phi \nabla ^2
\phi  - g_{\mu \nu } V(\phi )
\end{equation}
and
\begin{equation} \label{e18}
\nabla ^2 \phi  = \frac{2}{3}(- 3 + V(\phi)) \phi  + (1 -
\frac{1}{6} \phi^2) \frac{{dV}}{{d\phi }}
\end{equation}
with
\begin{equation} \label{e19}
V(\phi ) = \frac{1}{{12}}\phi ^4.
\end{equation}
Considering that all physical quantities are calculated in the new
frame, the metric and the scalar field of MTZ black hole are now
\begin{equation} \label{e20}
ds^2  =  - (r^2  - (1 + \frac{\mu }{r})^2 )dt^2  + (r^2 - (1 +
\frac{\mu }{r})^2 )^{ - 1} dr^2  + r^2 d\sigma ^2
\end{equation}
and
\begin{equation} \label{e21}
\phi  = \sqrt 6 \frac{\mu }{{r + \mu }}.
\end{equation}
Expressing the perturbation of scalar field $\tilde \phi = \phi +
\delta \phi$, we obtain the linear perturbation equation by
varying $\phi$ in Eq.(\ref{e18})
\begin{equation} \label{e22}
\nabla ^2 \delta \phi  = \delta \phi [\frac{1}{3}\phi
\frac{{dV}}{{d\phi }} + \frac{2}{3}(V - 3) + (1 - \frac{1}{6}\phi ^2
)\frac{{d^2 V}}{{d\phi ^2 }}].
\end{equation}

When we consider the perturbation of the scalar field $\tilde \phi=
\phi+ \delta \phi$, in principle the back-reaction of the metric
perturbation will affect the scalar field. However, the effect of
the metric fluctuation on the perturbation of the scalar can be
dropped near the critical point $\mu=0$. To see this, let us
consider a simple case where the perturbed metric is still of the
form (\ref{e20}) and  the metric function
$f=r^2-(1+\frac{\mu}{r})^2$ has a small fluctuation as $\tilde f = f
+ \delta f(t,r)$. Then linearizing the equation (\ref{e18}), we have
\begin{equation} \label{e38}
\nabla ^2 \delta \phi  + (\frac{{2\phi '}}{r} + \phi '')\delta f +
\phi '\delta f' = \delta \phi [\frac{1}{3}\phi \frac{{dV}}{{d\phi }}
+ \frac{2}{3}(V - 3) + (1 - \frac{1}{6}\phi ^2 )\frac{{d^2
V}}{{d\phi ^2 }}].
\end{equation}
 The last two terms in the left-hand-side of the
above equation are contributions from the back-reaction of the
metric perturbation. In our study, since our interest is focused on
the QNMs near the critical point, it is easy to show that the
factors of contributions from metric back-reaction tend to zero when
$\mu\rightarrow 0$ at the critical point by using (14). Thus the
back-reaction due to the metric perturbation can be neglected when
we focus on the behavior of the scalar perturbations around the
critical point $\mu=0$. Also near this point, the first and third
terms in the right hand side of the equation can be neglected. As a
result, near the critical point, the expression in the square
brackets in the right hand side of (\ref{e22}) turns to be $-2$.
Further, let us notice that the scalar field in the frame
(\ref{e20}) is a conformal scalar. This might explain that although
the two potentials shown in Fig.~1 look different, the numerical
results of the QNMs for two black holes are quite similar (see
Fig.~2 and 3). Of course, we may consider a scalar perturbation,
which is not related to the scaler field $\phi$ in the black hole
solution. In that case, we can obtain similar results for both black
holes.

By the variable separation $\delta \phi = \frac{1}{r} R(r)
Y(\Sigma) e^{-iwt} $, Eq.(\ref{e22}) is written as
\begin{eqnarray} \label{e23}
 R\left\{ {w^2  - V_{MTZ} } \right\} + R'ff' + R''f^2  = 0, \nonumber \\
 V_{MTZ}  =  (\xi ^2  + \frac{1}{4})\frac{f}{{r^2 }} - f(2 - \phi ^2 ) +
 \frac{{ff'}}{r}.
\end{eqnarray}
where $f = r^2 - (1 + \frac{\mu}{r})^2$ and $\xi^2 +1/4$ is the
eigenvalue of the harmonic function $Y(\Sigma)$ on the hyperbolic
space \cite{math}. Taking $R(r) = \tilde R(r) e^{-iw r_{*}}$ with
$dr_{*} = dr / f $, we can rewrite  Eq.(\ref{e23}) into
\begin{equation} \label{e24}
\tilde R\left\{ {-(\xi ^2  + \frac{1}{4})\frac{1}{{r^2 }} + (2 -
\phi ^2 ) - \frac{{f'}}{r}} \right\} + \tilde R'(f' - 2iw) + \tilde
R''f = 0.
\end{equation}
For the topological black hole Eq.(\ref{e8}), the scalar
perturbation is just described by the Klein-Gorden equation
\begin{equation} \label{e25}
\nabla ^2 \delta \phi  = 0
\end{equation}
which, under the variable separation $\delta \phi = \frac{1}{r}
R(r) Y(\Sigma) e^{-iwt} $ and the transformation $R(r) = \tilde
R(r) e^{-iw r_{*}}$ with $dr_{*} = dr / g $, it can be expressed
as
\begin{eqnarray} \label{e26}
  - \frac{{V_{TBH} }}{f_1}\tilde R + \tilde R'(f_1' - 2iw) + \tilde R''f_1 = 0, \nonumber \\
 V_{TBH}  =  f_1[(\xi ^2  + \frac{1}{4})\frac{1}{{r^2 }} + \frac{{f_1'}}{r}].
\end{eqnarray}
Here the metric coefficient $f_1 = r^2 - 1 - \frac{{2 \mu}}{r}$.

The behaviors of effective potentials of scalar perturbations in
the MTZ and topological black holes' backgrounds are shown in
Fig.\ref{f1}. Compared with the situation in the topological black
hole, the effective potential in the MTZ background does not
diverge, but converges to a nonzero constant at spatial infinity,
due to the compensation of the divergence of the non-zero scalar
field in Eq.(\ref{e23}). For the convergent potential at spatial
infinity, Koutsoumbas {\it et al.} \cite{14} imposed the boundary
condition that the wave function of the electromagnetic
perturbation vanishes at spatial infinity and employed the
numerical method devised by Horowitz and Hubeny \cite{4} in the
MTZ background. In our case, we can see that the tortoise
coordinate $r_*$ tends to a constant for big $r$ in the MTZ
background, which is similar to that of the Schwarzschild-AdS
observed in \cite{PP}. This leads $e^{-i w r_*}$ to be a constant
for big $r$, which indicates that there is no flux at spacial
infinity. As for $\tilde R(r)$, we can obtain the approximate
solution $\tilde R \sim e^{x \cdot Const.}$ after substituting $r$
with $1/x$ in Eq.(\ref{e24}) and taking the leading terms near $x
= 0$. Therefore $\tilde R $ is also a constant at the spacial
infinity $x=0$. In other words, the oscillation of the QNMs is
frozen, hence the wave vanishes at the infinity and this constant
can be set zero by the definition of QNM. This makes us confident
to employ the method of Horowitz and Hubeny to calculate the QNMs
of the MTZ black hole. For the topological black hole case,
Horowitz and Hubeny method applies naturally.

\begin{figure}[ht]
\vspace*{0cm}
\begin{minipage}{0.3\textwidth}
\resizebox{1.1\linewidth}{!}{\includegraphics*{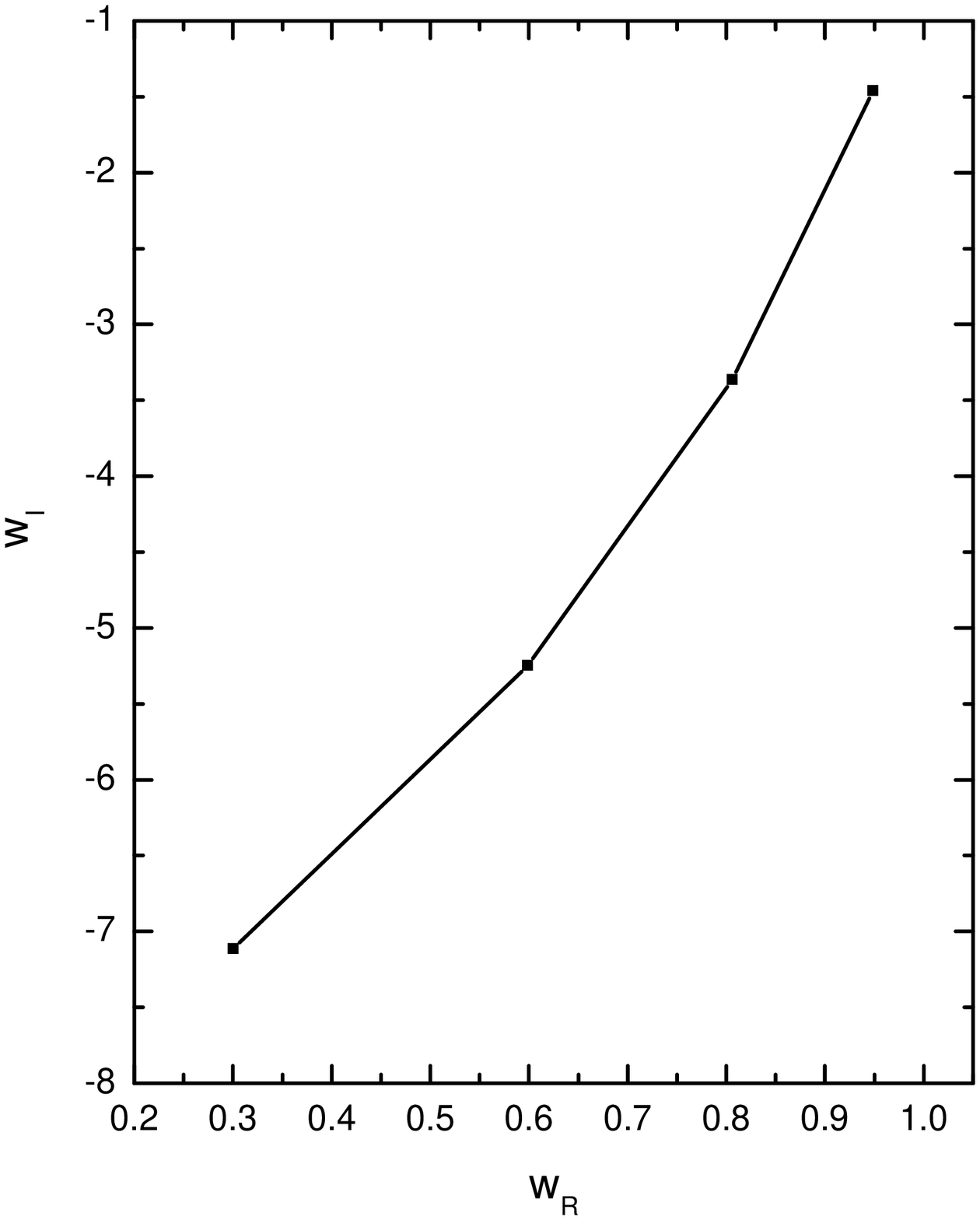}}
\nonumber
\end{minipage}\nonumber
\begin{minipage}{0.3\textwidth}
\vspace*{0.0cm}
\resizebox{1.1\linewidth}{!}{\includegraphics*{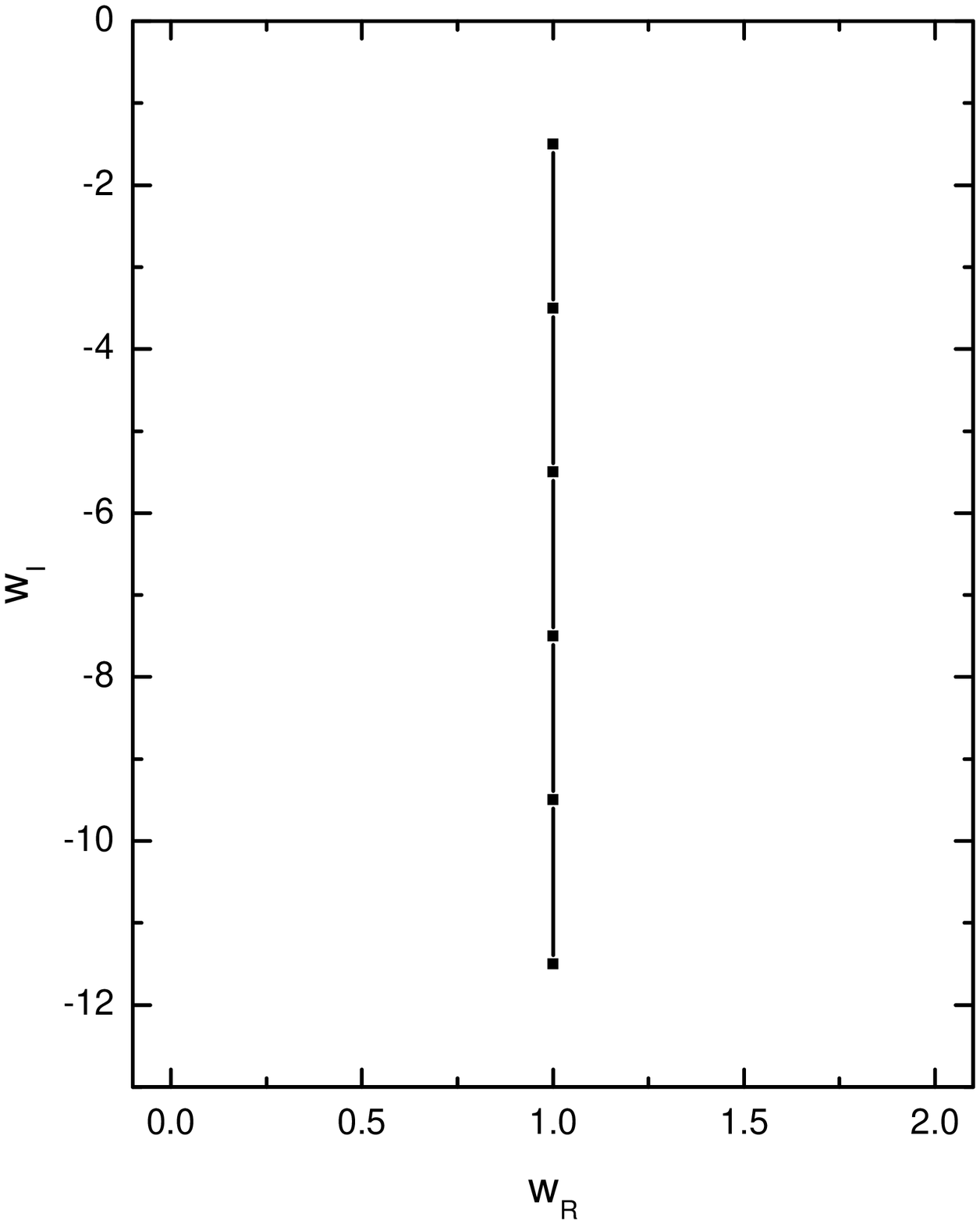}}
\nonumber
\end{minipage} \nonumber
\begin{minipage}{0.3\textwidth}
\vspace*{0.0cm}
\resizebox{1.1\linewidth}{!}{\includegraphics*{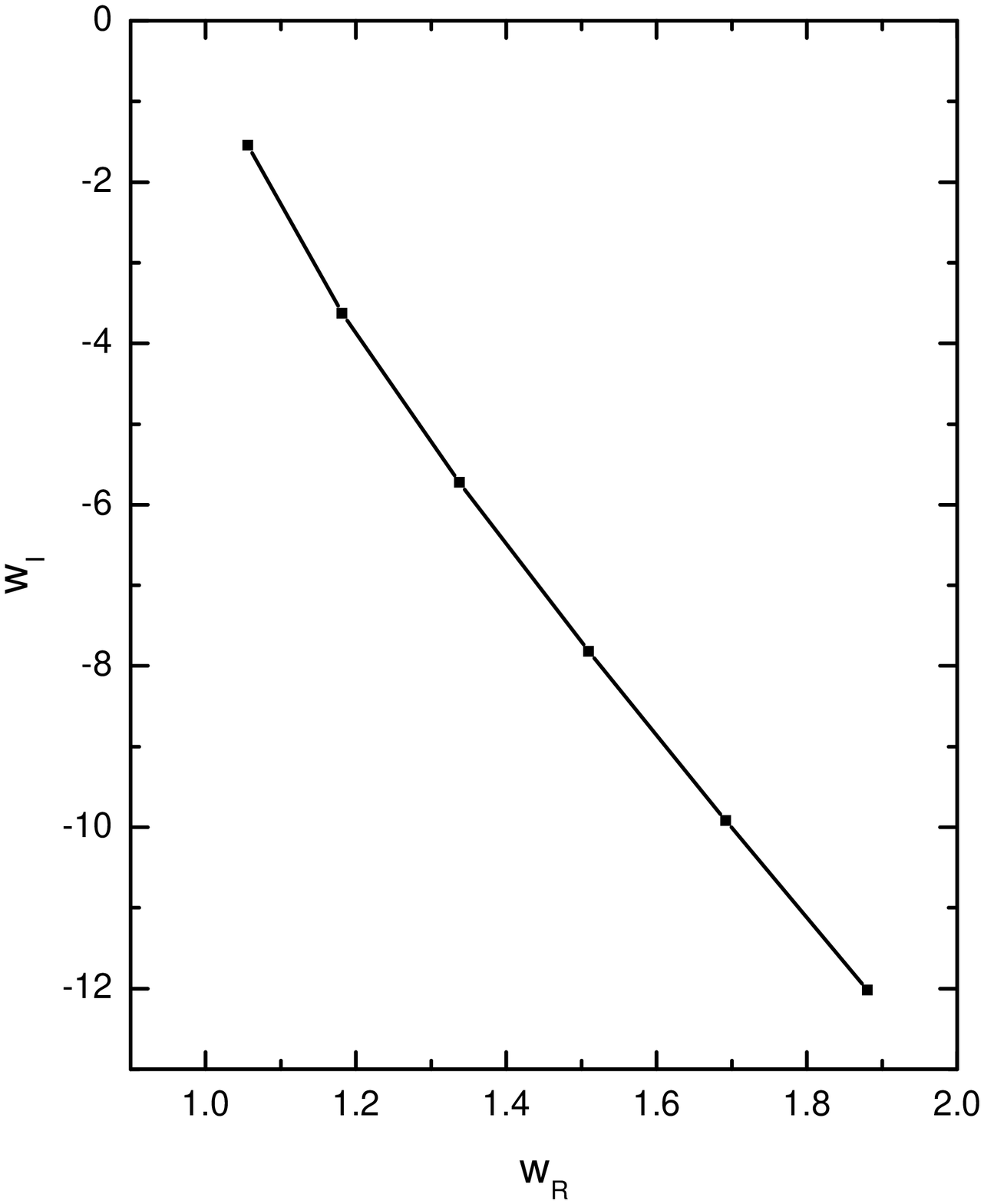}}
\nonumber
\end{minipage} \nonumber
\caption{{The QNMs of scalar perturbations in the MTZ black hole.
The results are calculated with $\xi=1$ in the Eq.(\ref{e24}). The
right, middle and left figures stand for the cases of $r_+ = 0.97$,
$1.00$ and $1.03$, respectively.}} \label{f2}
\end{figure}

\begin{figure}[ht]
\vspace*{0cm}
\begin{minipage}{0.3\textwidth}
\resizebox{1.1\linewidth}{!}{\includegraphics*{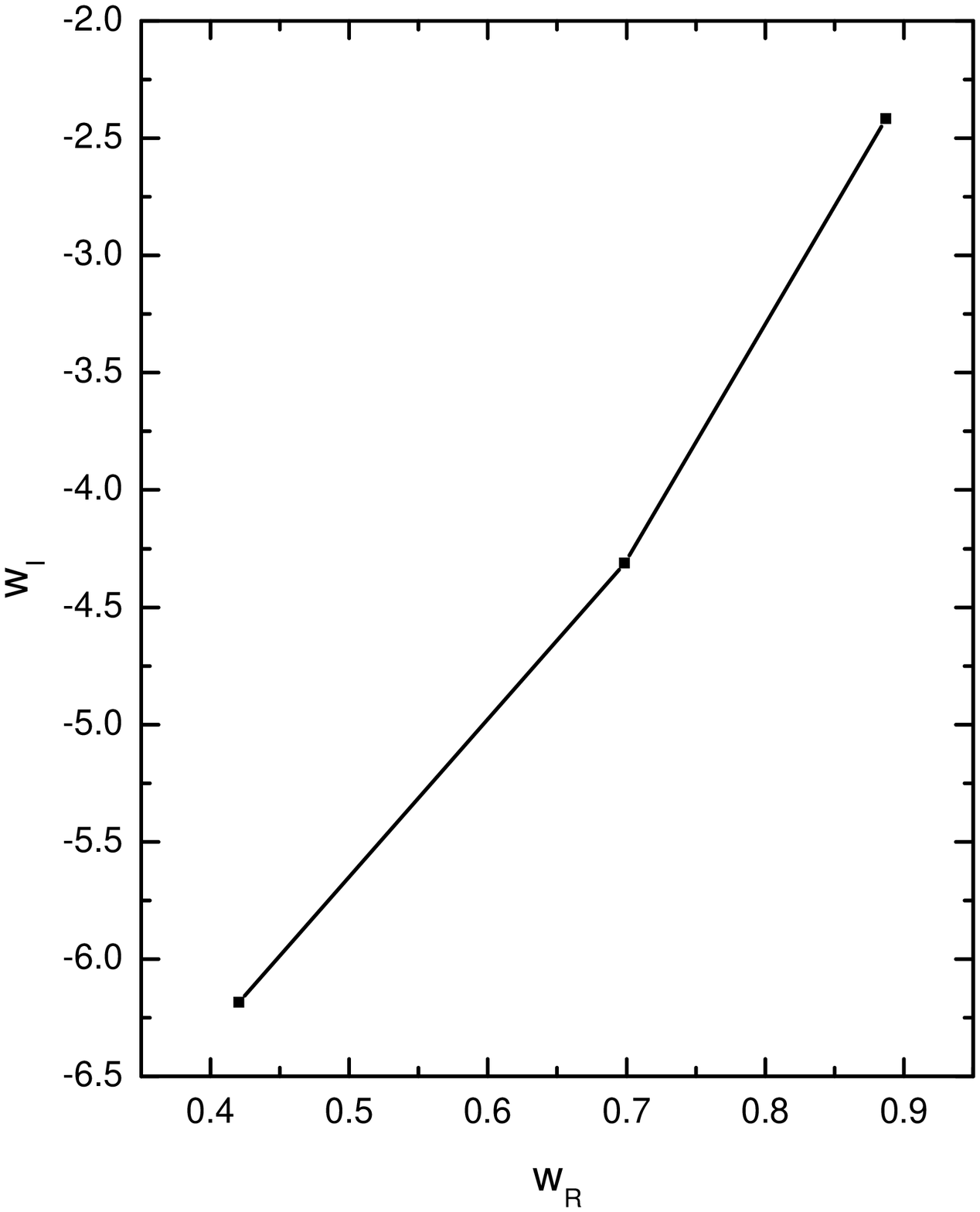}}
\nonumber
\end{minipage}\nonumber
\begin{minipage}{0.3\textwidth}
\vspace*{0.0cm}
\resizebox{1.1\linewidth}{!}{\includegraphics*{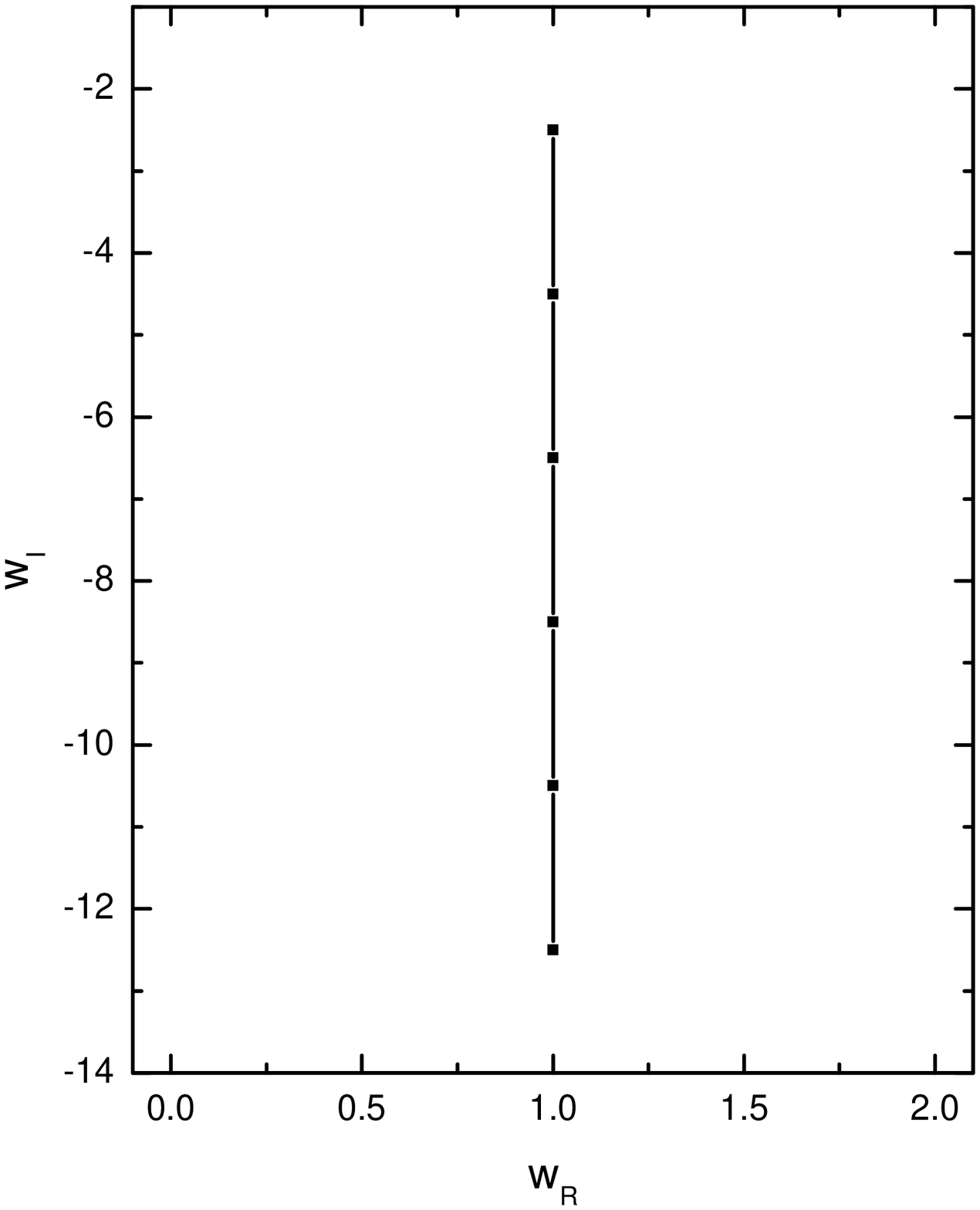}}
\nonumber
\end{minipage} \nonumber
\begin{minipage}{0.3\textwidth}
\vspace*{0.0cm}
\resizebox{1.1\linewidth}{!}{\includegraphics*{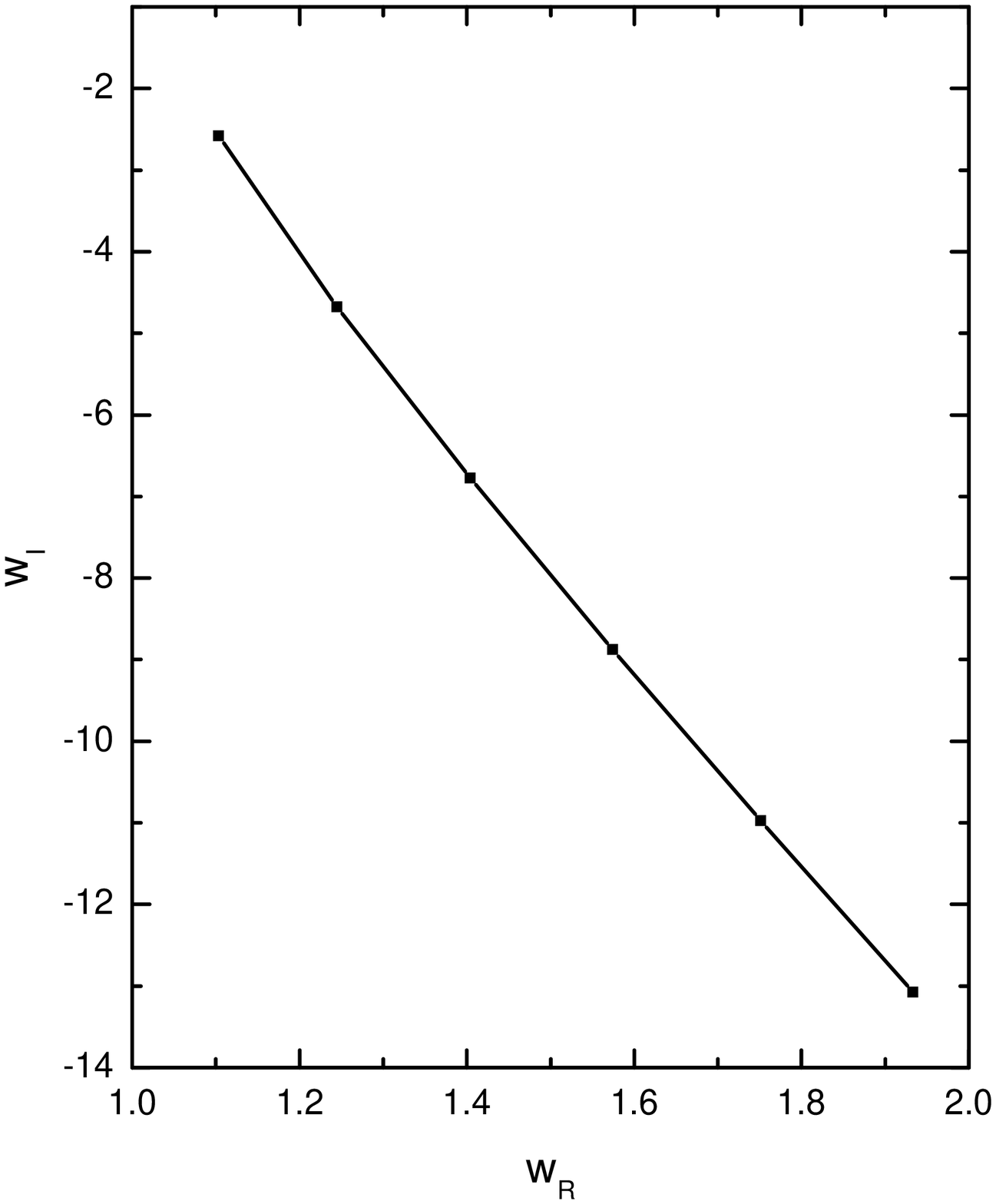}}
\nonumber
\end{minipage} \nonumber
\caption{{The QNMs of scalar perturbation in the topological black
hole. The results are calculated with $\xi=1$ in the Eq.(\ref{e26}).
The right, middle and left figures stand for the cases of $r_+ 0.97$, $1.00$ and $1.03$, respectively.}} \label{f3}
\end{figure}

In the following we present our numerical results of  QNMs for the
MTZ and topological black holes when their event horizons cross
the critical point $r_c =1$, e.g. $r_+ = 0.97, 1, 1.03$. Our
results of  are shown in Fig.\ref{f2} and Fig.\ref{f3}
respectively.

We can see that when $r_+ < 1$, the slope of QNMs is positive and
the slope tends to minus infinity as we approach $r_+=1$. When
$r_+>1$, the QNMs lie on a straight line with negative slope. The
results of the QNMs of scalar perturbation present us very similar
properties to those in the electromagnetic perturbations
\cite{14}. As argued in the study of the electromagnetic
perturbation, the change of slope of the QNMs in the scalar
perturbations as we decrease the value of the horizon radius below
a critical value reveals the phase transition of a vacuum
topological black hole to the MTZ hole with scalar hair. Our
result presents a support to \cite{14} in the study of the phase
transition for the four-dimensional topological black holes with
scalar hair.

\section{AdS black holes with Ricci flat horizons on the AdS soliton background}

To investigate whether the QNMs is an effective tool to disclose
phase transition in general black hole configurations, we are going
to extend the discussion to phase transitions for flat AdS black
holes in this section. Thermodynamics of AdS black holes with Ricci
flat horizons using the AdS soliton as the thermal background has
been investigated in \cite{HM, surya}. It was found that there is a
phase transition analogous to the Hawking-Page transition. We will
compute the QNMs of the scalar perturbation in the flat AdS black
hole and AdS soliton backgrounds and examine whether the phase
transition imprints in the QNMs.

The metric of the AdS black hole with Ricci flat horizon (flat AdS
black hole) is \cite{surya}
\begin{equation} \label{e27}
ds_{bh} ^2  =  - V_b dt_b^2  + V_b ^{ - 1} dr^2  + r^2 d\phi _b ^2
+ r^2 h_{ij} d\theta _i d\theta _j,
\end{equation}
where
\begin{equation} \label{e28}
V_b  = r^2  - \frac{{k_b^{n - 1} }}{{r^{n - 3} }}
\end{equation}
with $l=1$ for simplicity. $h_{ij}$ is a Ricci flat metric and
$\phi_b$ is identified with period $\eta_b$. $k_b$ is the black hole
mass parameter. The horizon is at $V_b(r_{b+})=0$ and the zeros of
$V_b(r)$ are given by $r^{n-1}_{b+}=k^{n-1}_{b}$. By the analytical
continuation $t_b \to i \phi_s$ and $\phi_b \to t_s$, the AdS
soliton is got
\begin{equation} \label{e29}
ds_s ^2  =  - r^2 dt_s ^2  + V_s ^{ - 1} dr^2  + V_s d\phi _s ^2  +
r^2 h_{ij} d\theta _i d\theta _j,
\end{equation}
where $V_s$ is the function Eq.(\ref{e28}) with $k_s$ replacing
$k_b$. To meet the requirement of regularity, $r \geq r_{s+} k_s$, where $V_s(r_{s+})$ vanishes and $\phi_s$ is identified with
the period $\beta_s = \frac{{4 \pi}}{{3 r_{s+}}}$.

To calculate the energy and to study their thermodynamics, the
standard regularization scheme is used\cite{surya}, in which these
two solutions are matched at a finite cutoff radius $R$
\begin{equation} \label{e30}
\beta _b \sqrt {V_b }  = R\eta _s ,\quad \beta _s \sqrt {V_s }  R\eta _b
\end{equation}
and then the limit $R \to \infty$ is taken after all quantities
are calculated. By comparing the energies and the Euclidean
actions of the two solutions, Surya et.al. \cite{surya} suggested
a phase transition for $n > 3$ analogous to the Hawking-Page
transition. If $k_b \ll k_s$, small hot black hole is unstable and
decays to small hot soliton, while large cold black hole is stable
with $k_b \gg k_s$. In case of $k_b \sim k_s$, black hole is in
equilibrium with the soliton including cases when they are large
and hot, or cold and small.

\begin{figure}[ht]
\vspace*{0cm}
\begin{minipage}{0.3\textwidth}
\resizebox{1.1\linewidth}{!}{\includegraphics*{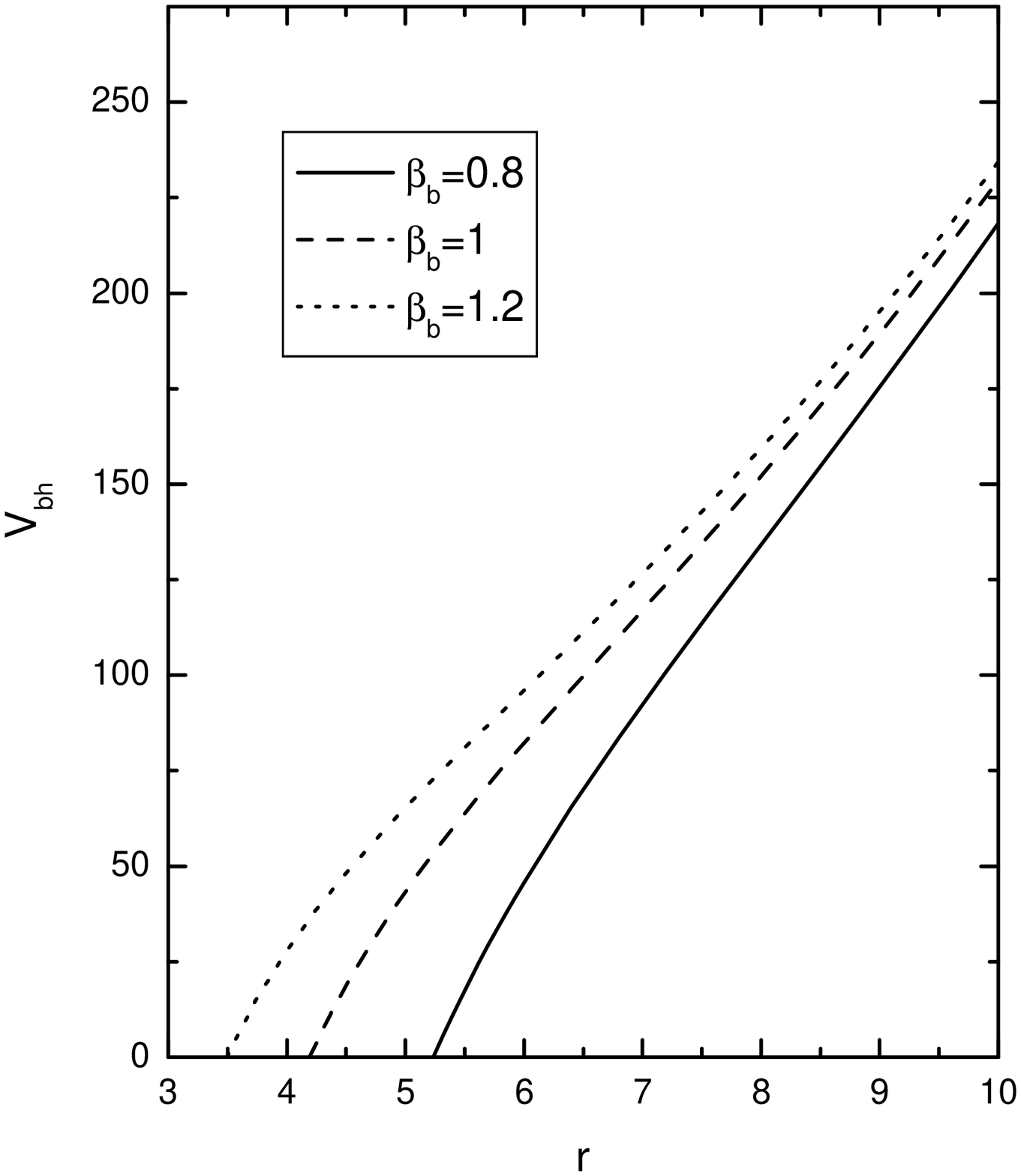}} \nonumber
\end{minipage}\nonumber
\begin{minipage}{0.3\textwidth}
\vspace*{0.0cm}
\resizebox{1.1\linewidth}{!}{\includegraphics*{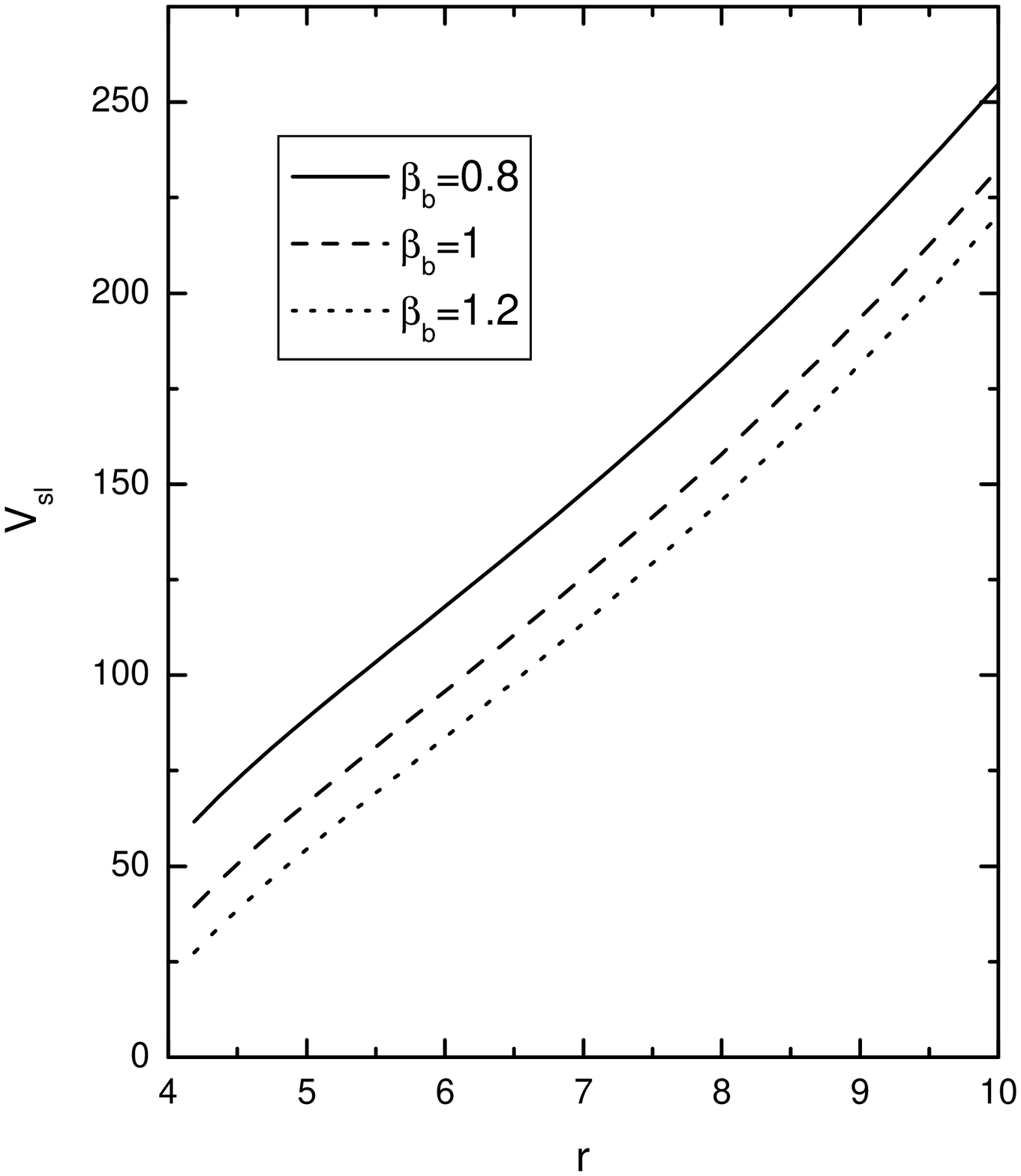}} \nonumber
\end{minipage} \nonumber
\caption{{The effective potentials of the scalar perturbations. The
left one stands for the flat AdS black hole and the right for the
AdS soliton, when $k=1$, $n=1$ and $\beta_s = 1$ fixed in the
Eq.(\ref{e34}) and Eq.(\ref{e36}).}} \label{f4}
\end{figure}

In order to see the signature of this phase transition in QNMs, we
study the scalar perturbations in these two solutions with the
dimension $n=4$. By separating variables $\psi_b R_b(r,t)\Phi_b(\phi_b)Y(\theta)$ in the KG equation $\nabla^2
\psi_b = 0$ of the flat AdS black hole Eq.(\ref{e27}), we have
\begin{equation} \label{e31}
\frac{{R_b }}{{r^2 }}\left\{ {\frac{1}{{\Phi _b }}\frac{{d^2 \Phi _b
}}{{d\phi _b ^2 }} + \frac{1}{Y}\frac{{d^2 Y}}{{d\theta ^2 }}}
\right\} + (\frac{{2V_b }}{r} + V_b ')\frac{{\partial R_b
}}{{\partial r}} + V_b \frac{{\partial ^2 R_b }}{{\partial r^2 }} -
\frac{1}{{V_b }}\frac{{\partial ^2 R_b }}{{\partial t^2 }} = 0 .
\end{equation}
The second term of the square bracket just stands for the free
motion in the $\theta$ direction and then its eigenvalue is the
kinetic energy $- k^2$ with any real number $k$. The first term also
can be considered as the free motion in the periodic coordinate
$\phi_b$, therefore the eigenvalue is not a real number but discrete
$- \sigma_b^2 = - (\frac{{2 \pi n}}{\eta_b} )^2$ $(n=0,1,2...)$.
Taking Eq.(\ref{e30}) at the infinity $R \to \infty$
\begin{equation} \label{e32}
\sigma _b^2  = \left( {\frac{{2\pi n}}{{\eta _b }}} \right)^2  \left( {\frac{{2\pi n}}{{\beta _s }}} \right)^2,
\end{equation}
Eq.(\ref{e31}) can be written as
\begin{equation} \label{e33}
 - \frac{{R_b }}{{r^2 }}(\sigma _b^2  + k^2 ) + (\frac{{2V_b }}{r} + V_b ')\frac{{\partial R_b }}{{\partial r}} + V_b \frac{{\partial ^2 R_b }}{{\partial r^2 }} - \frac{1}{{V_b }}\frac{{\partial ^2 R_b }}{{\partial t^2 }}  0.
\end{equation}
Setting $R_b(t,r) = \frac{1}{r} \tilde R_b(r) e^{-i w t}$, the
above equation changes into
\begin{eqnarray} \label{e34}
  - \tilde R_b \left\{ { - w^2  + V_{bh} } \right\} + V_b 'V_b \tilde R_b ' + V_b ^2 \tilde R_b '' = 0, \nonumber \\
 V_{bh}  = \frac{{2V_b V_b '}}{r} + \frac{{V_b }}{{r^2 }}(\sigma _b^2  + k^2
 ).
\end{eqnarray}
Near the horizon $r_{b+} = k_b$, the wave can be expanded as
$\tilde R_b(r) = \sum a^{(b)}_i (r- r_{b+})^{\rho + i} $ and the
index $\rho = \pm \frac{{i w}}{{3 r_{b+}}}$ through the index
equation. The effective potential $V_{bh}$ is shown in
Fig.\ref{f4} which diverges at the spacial infinity. Because there
only exists the ingoing wave near the horizon, the positive sign
in $\rho$ is discarded.

In the case of the AdS soliton Eq.(\ref{e29}), we follow the
similar steps to that of the flat AdS black hole with the variable
separation $\psi_s = R_s(r,t)\Phi_s(\phi_s)Y(\theta)$ in KG
equation and obtain
\begin{equation} \label{e35}
R_s \left\{ {\frac{1}{{V_s }}\frac{1}{{\Phi _s }}\frac{{d^2 \Phi _s
}}{{d\phi _s ^2 }} + \frac{1}{{r^2 }}\frac{1}{Y}\frac{{d^2
Y}}{{d\theta ^2 }}} \right\} + (\frac{{2V_s }}{r} + V_s
')\frac{{\partial R_s }}{{\partial r}} + V_s \frac{{\partial ^2 R_s
}}{{\partial r^2 }} - \frac{1}{{r^2 }}\frac{{\partial ^2 R_s
}}{{\partial t^2 }} = 0.
\end{equation}
As above, the eigenvalues of the second and first terms are $-k^2$
and $ - \sigma_s^2 = - (\frac{{2 \pi n}}{\eta_s} )^2$ $(n=0,1,2...)$
respectively. Hence the radial part of KG equation in the AdS
soliton background is
\begin{eqnarray} \label{e36}
  - \tilde R_s \left\{ { - \frac{{w^2 V_s }}{{r^2 }} + V_{sl} } \right\} + V_s 'V_s \tilde R_s ' + V_s ^2 \tilde R_s '' = 0, \nonumber \\
 V_{sl}  = \frac{{2V_s 'V_s }}{r} + \sigma _s^2  + \frac{{k^2 V_s }}{{r^2 }}
\end{eqnarray}
with $R_s(t,r) = \frac{1}{r} \tilde R_s(r) e^{-i w t}$ and
\begin{equation} \label{e37}
\sigma _s^2  = \left( {\frac{{2\pi n}}{{\eta _s }}} \right)^2  \left( {\frac{{2\pi n}}{{\beta _b }}} \right)^2.
\end{equation}
Again, the index $\rho$ in the expansion $\tilde R_s(r) = \sum
a^{(s)}_i (r- r_{s+})^{\rho + i} $ is given by $\rho = \pm
\frac{{\sigma_s}}{{3 r_{b+}}}$. The negative sign is discarded to
keep the convergence of the wave near the horizon $r_{s+}$. We plot
$V_{sl}$ in Fig.\ref{f4} as the comparison to $V_{bh}$. Notice that
the real index $\rho$ is the main difference from that of the flat
AdS black hole which indicates there is no wave flux near the
boundary $r_{s+}$. In the far region, the AdS soliton approaches the
AdS spacetime and hence no wave can flow to the infinity. Therefore,
only normal modes exist similar to the form of stationary waves in
the AdS soliton background, once some perturbations are excited. Our
following numerical calculation proves this effect.

\begin{figure}[ht]
\vspace*{0cm}
\begin{minipage}{0.3\textwidth}
\resizebox{1.1\linewidth}{!}{\includegraphics*{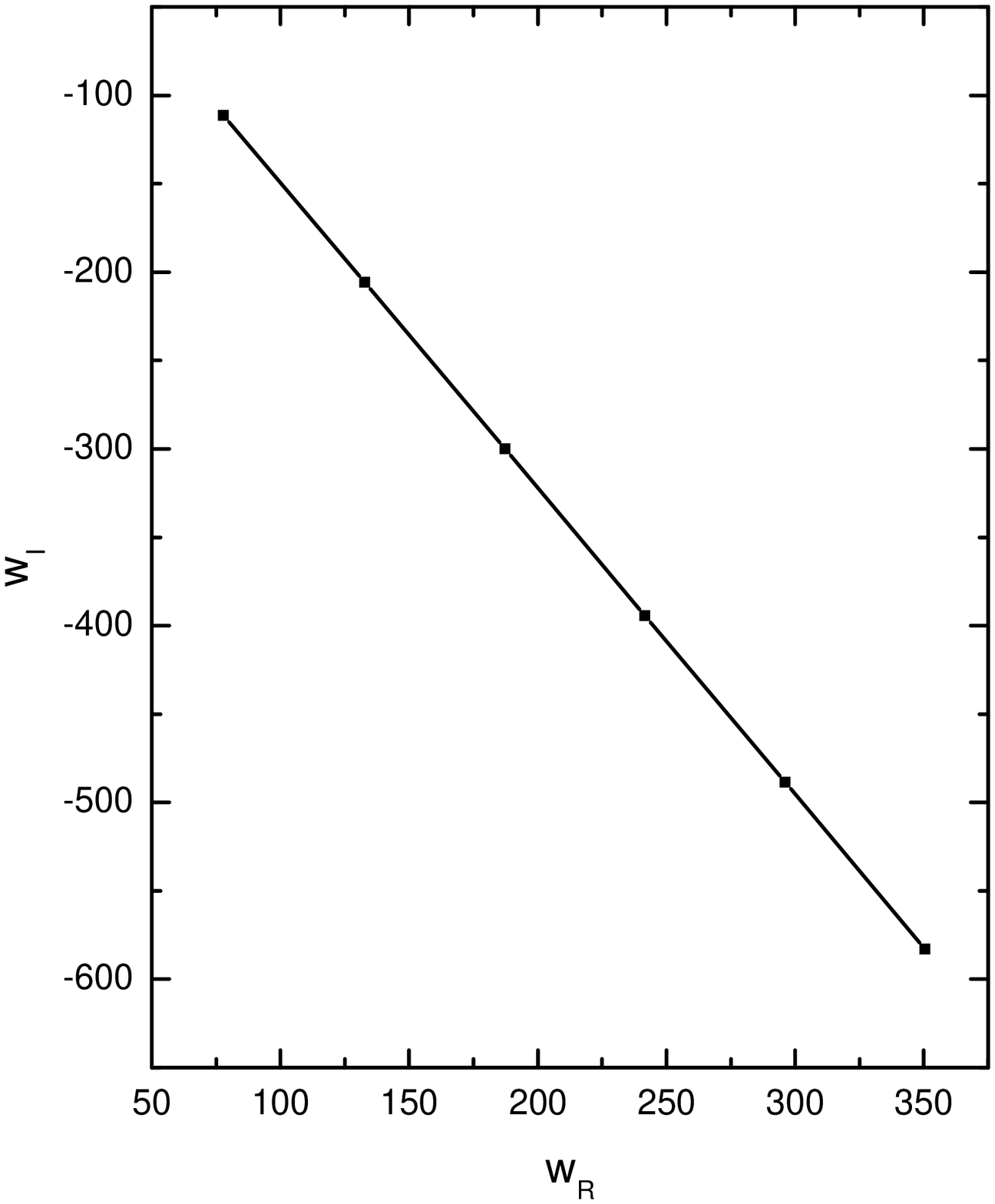}} \nonumber
\end{minipage}\nonumber
\begin{minipage}{0.3\textwidth}
\vspace*{0.0cm}
\resizebox{1.1\linewidth}{!}{\includegraphics*{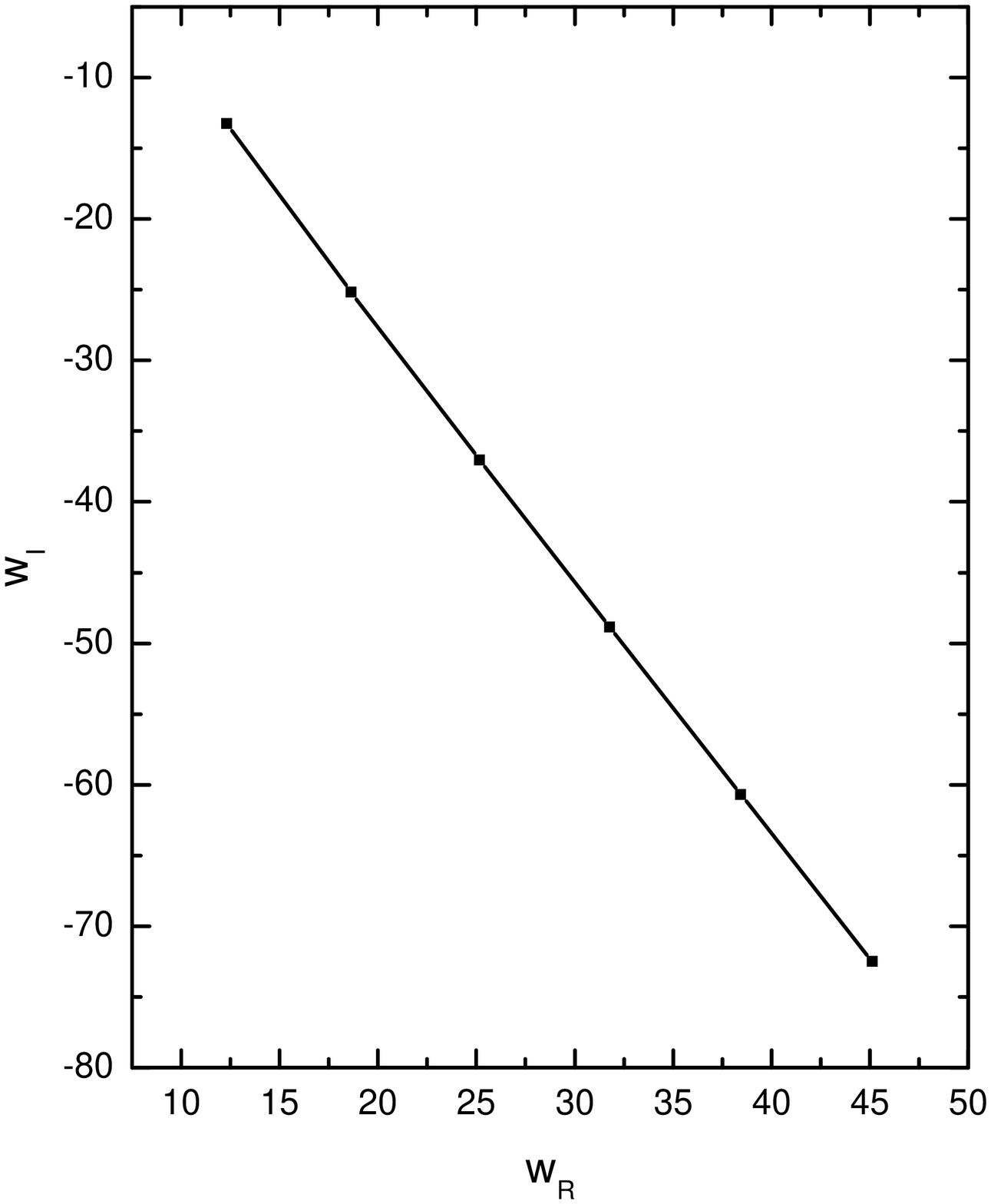}} \nonumber
\end{minipage} \nonumber
\begin{minipage}{0.3\textwidth}
\vspace*{0.0cm}
\resizebox{1.1\linewidth}{!}{\includegraphics*{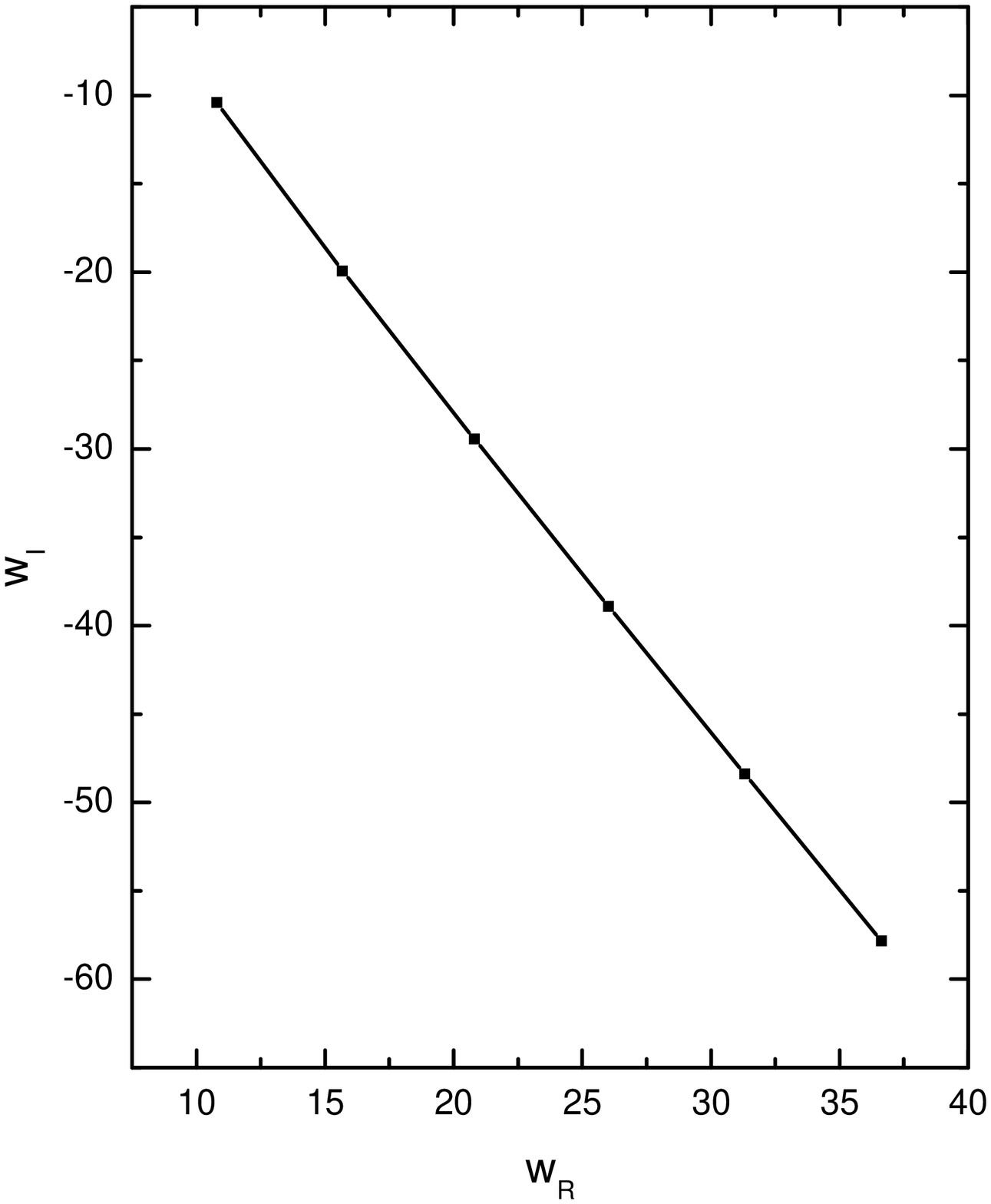}} \nonumber
\end{minipage} \nonumber
\begin{minipage}{0.3\textwidth}
\vspace*{0.0cm}
\resizebox{1.1\linewidth}{!}{\includegraphics*{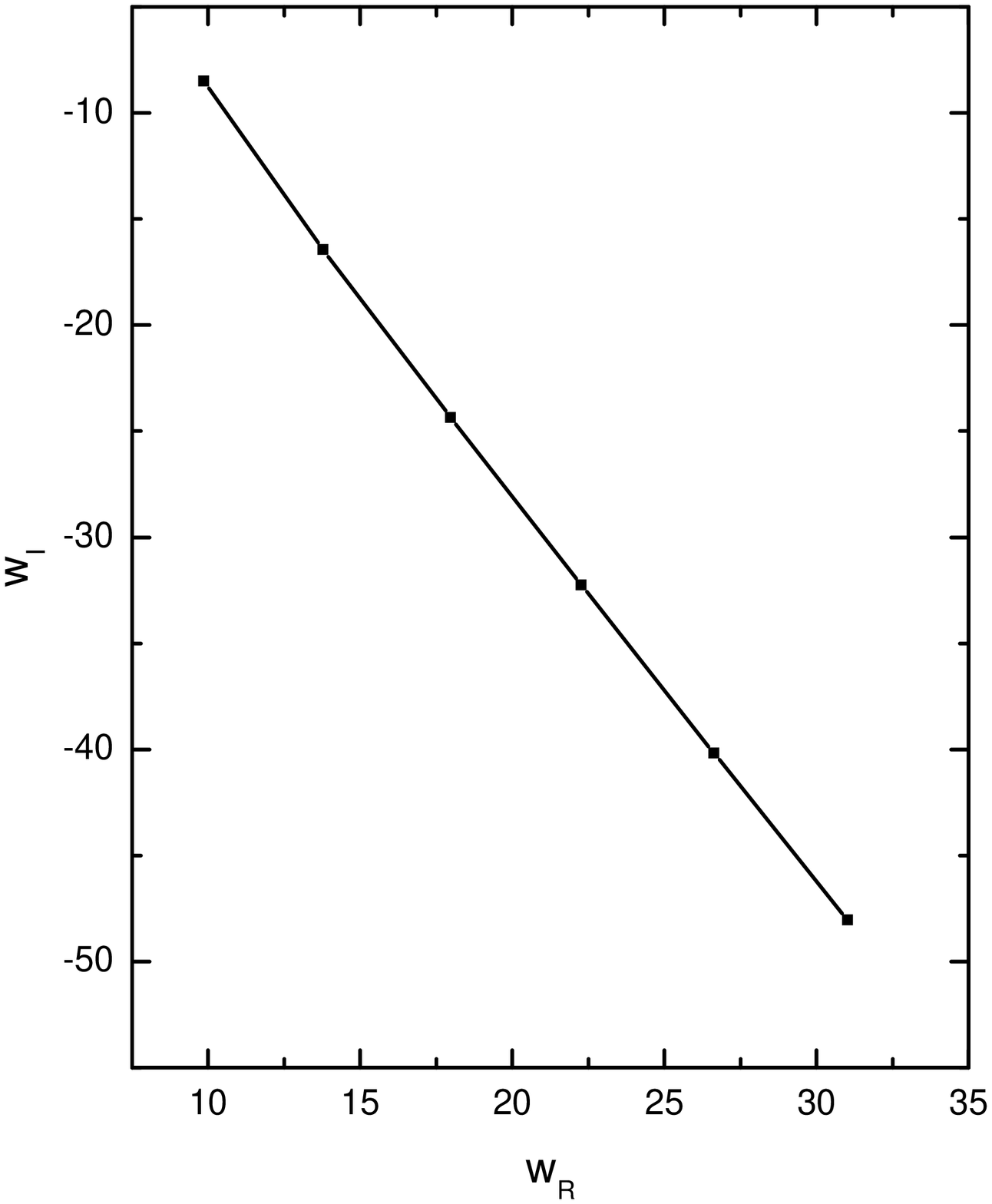}} \nonumber
\end{minipage} \nonumber
\begin{minipage}{0.3\textwidth}
\vspace*{0.0cm}
\resizebox{1.1\linewidth}{!}{\includegraphics*{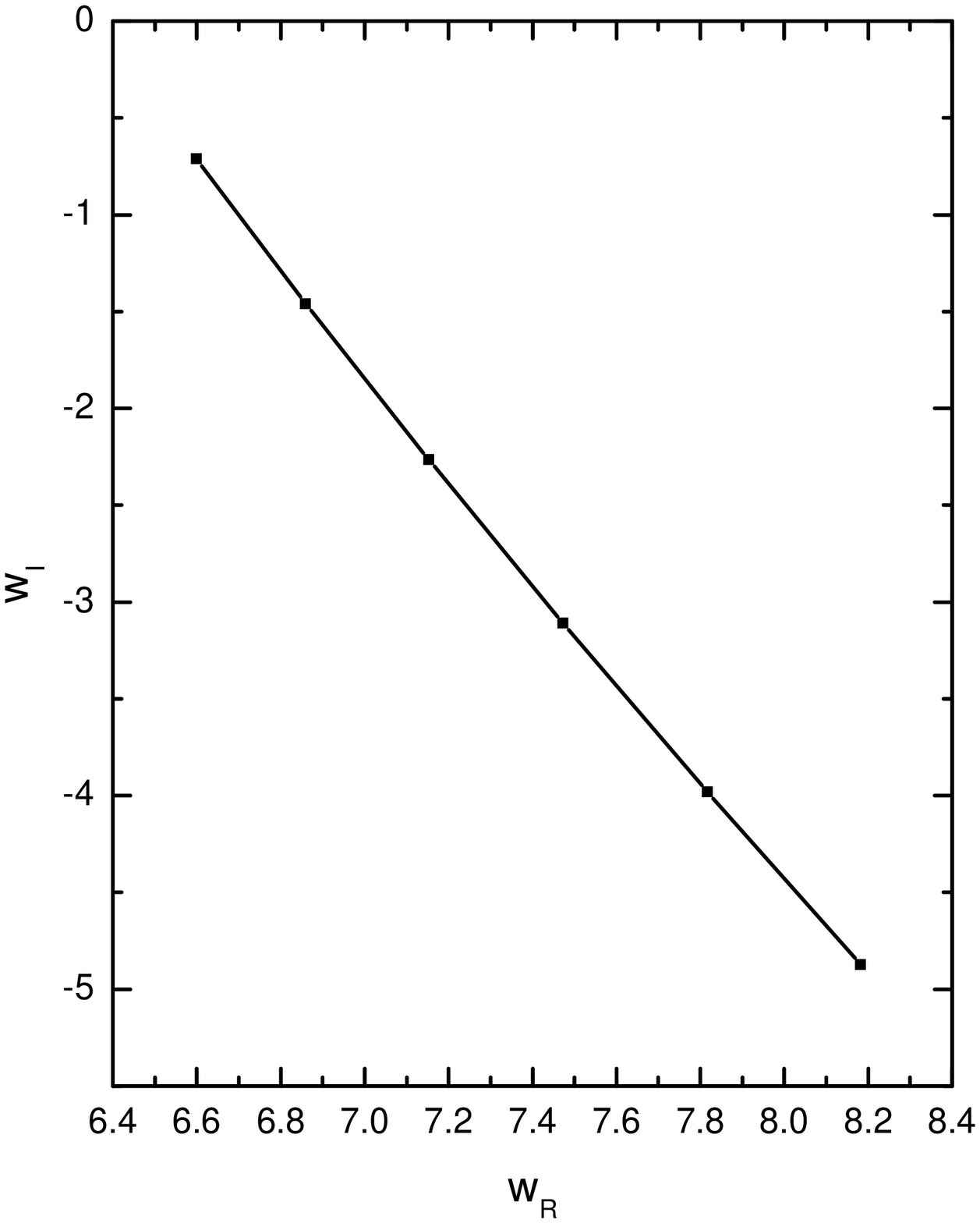}}
\nonumber
\end{minipage} \nonumber
\caption{{The QNMs of scalar perturbation in the flat AdS black
hole. The results are calculated with $k = 1$ and $n=1$ in
Eq.(\ref{e34}). The right, middle and left figures in the first
row stand for the cases of $\beta_b = 0.1$, $0.8$ and $1.0$
respectively and the right and left ones in the second row for
$\beta_b = 1.2$ and $10$}} \label{f5}
\end{figure}

We still employ the method of Horowitz and Hubney to calculate the
numerical QNMs of the flat AdS black hole and AdS soliton. We do
the computation with $k = 1$ and $n=1$ in Eq.(\ref{e34}) and
Eq.(\ref{e36}), because the free motion in the $\theta$ direction
is not of importance and it is better that the wave number $n$ of
the motion in the $\phi_{\alpha}$ $( \alpha = b,s )$ direction
does not vanish if we consider the more differences between the
flat AdS black hole and AdS soliton in the QNMs. During the
computation, we fix the period $\beta_s = 1.0$ and vary $\beta_b$
from 0.1 to 10 representing the cases where $\beta_b$ is much
smaller to much bigger than $\beta_s$ in order to satisfy the
condition of the phase transition. Fig.\ref{f5} and Table.I show
the QNMs of the scalar perturbation in the two spacetimes. The
QNMs in the flat AdS black hole always have negative slope in the
$w_R - w_I$ diagrams, distinguished from the situation of the
MTZ-TBH phase transition. As for the AdS soliton background, there
only exist the normal modes as we expected and no special events
happen here. We can see that flat AdS black hole and AdS soliton
are staying in different phases as disclosed by the QNMs behavior,
however the phase transition as revealed in \cite{surya} occuring
around the critical point $\beta_b \sim \beta_s = 1$ does not
imprint explicitly in the present QNMs study. More detailed
analysis is called for to reveal more subtle changes in the QNMs
due to this phase transition.

\vspace{0.5cm}
\begin{center}

The Normal Modes $\omega$

\begin{tabular}{l|l|l|l|l|l|l} \hline
\hline
$\beta_b=0.1$ &   $118.869$   &  $214.033$ &   $308.428$   &  $402.572$ &   $496.607$   &  $590.583$ \\
$\beta_b=0.8$  &   $19.381$   &  $31.108$ &   $42.835$   &  $54.563$ &   $66.293$   &  $78.022$ \\
$\beta_b=1$  &   $16.590$   &  $25.921$ &   $35.278$   &  $44.646$ &   $54.019$   &  $63.396$ \\
$\beta_b=1.2$  &   $14.743$   &  $22.475$ &   $30.250$   &  $38.042$ &   $45.844$   &  $53.652$ \\
$\beta_b=10$  &   $7.097$   &  $7.855$ &   $8.662$   &  $9.498$ &   $10.356$   &  $11.228$ \\
\hline
\end{tabular}

\vspace{0.1cm} \large{Table I}

\vspace{0.1cm} {\it The normal modes of the scalar field in the AdS
soliton background with fixed $\beta_s = 1$. }

\end{center}
\vspace{0.5cm}

\section{Conclusions and discussions}
Because of its astrophysical and theoretical interests, the QNMs
of black holes has been an intriguing subject of discussions.
Calculating the QNMs of electromagnetic perturbations of the MTZ
and topological black holes, Koutsoumbas et al\cite{14} argued
that the QNMs imprints the phase transition of a vacuum
topological black hole to the MTZ black hole with scalar hair. To
examine whether this interesting result is just an accident, we
have studied the QNMs for the general scalar perturbations. We
observed in the numerical investigation that the slope of the QNMs
changes as we decrease the value of the horizon radius below a
critical value, which is in agreement with the behavior observed
for the electromagnetic perturbations\cite{14}. Thus for the
four-dimensional topological black holes with scalar hair, the
QNMs really presents the signature of the phase transition.

It would be fair to say that we just have the thermodynamical
descriptions of the black holes' phase transitions, the physical
phenomenons of the phase transitions are not clear. If the QNMs
can probe the phase transition as disclosed in topological black
holes with scalar hair, it would be interesting to ask whether
this tool is effective for general black hole configurations. We
have investigated the QNMs of the scalar perturbations in the
backgrounds of flat AdS black holes and AdS solitons. Although it
is clear from the QNMs that flat AdS black holes and AdS solitons
are in different phases, it is not clear as observed in MTZ and
topological black holes that near the critical point there is a
sudden change in the QNMs behavior. More detailed careful study of
QNMs is called for to disclose more subtle changes in the QNMs
caused by the phase transition.

\begin{acknowledgments}
This work was partially supported by NNSF of China, Ministry of
Education of China and Shanghai Education Commission. JY Shen was
supported by the Fudan Creative Fund for Graduate Student.

\end{acknowledgments}

\end{document}